\pdfoutput=1

\documentclass{article}

\usepackage[a4paper,left=4cm]{geometry}	% Page layout
\usepackage{lmodern,microtype}	% Allowing modern fonts and more flexibility
\usepackage{tikz}				% For pictures
\usepackage[english]{babel}		% Language
\usepackage{color}				% Colours
\usepackage{enumitem}
\usepackage{amssymb,amsthm}
\usepackage{amsmath}
\usepackage{bm}
\usepackage{authblk}
\usepackage{array}
\usepackage{stackrel}

\definecolor{darkblue}{rgb}{0,0,.8}

\usepackage[nobreak]{cite} 

\usepackage{hyperref}
\hypersetup{
    colorlinks,
    citecolor=red,
    filecolor=black,
    linkcolor=darkblue,
    urlcolor=black
}

\usepackage[capitalise,noabbrev]{cleveref}

\newcommand{\chit}{\protect\raisebox{0.25ex}{$\chi$}}
\renewcommand{\i}{\text{i}}

\newcommand{\q}{\mathfrak{q}}
\newcommand{\C}{\mathbb{C}}
\newcommand{\id}{\mathbb{I}}
\newcommand{\ot}{\otimes}
\newcommand{\sli}{\sum\limits}
\newcommand{\ra}{\rightarrow}
\newcommand{\gl}[2]{\ensuremath{\mathfrak{gl}\left({#1}|{#2}\right)}}

\newcommand{\be}{\begin{eqnarray}}
  \newcommand{\ee}{\end{eqnarray}}
\newcommand{\ben}{\begin{eqnarray*}}
  \newcommand{\een}{\end{eqnarray*}}
\newcommand{\bec}{\begin{equation}\begin{array}{lll}}
    \newcommand{\eec}{\end{array}\end{equation}}

\title{Lattice SUSY for the DiSSEP at $\lambda^2=1$ (and $\lambda^2 = - 3 $)}

\author{Desmond A. Johnston\thanks{\tt D.A.Johnston@hw.ac.uk}\;}
\affil{School of  Mathematical and Computer Sciences,\\ Heriot Watt University,\\Edinburgh EH14 4AS, UK}
\begin{document}
\maketitle

\bibliographystyle{unsrt}
\begin{abstract}
  \noindent 
We investigate whether  the dynamical lattice supersymmetry  discussed for various Hamiltonians, including one-dimensional quantum spin chains,  by Fendley et.al. \cite{Fend1,Fend2,Fend3} and Hagendorf  et.al. \cite{Hag1,Hag2,Hag3} might  also exist for the Markov matrices of  any one-dimensional exclusion processes, since these  can be related by conjugation to quantum spin chain Hamiltonians.

We find that the 
DiSSEP (Dissipative  Symmetric Simple Exclusion Process), introduced by Crampe et.al. in \cite{DiSSEP,these},  provides one such example for suitably chosen parameters. The DiSSEP Markov matrix admits the supersymmetry  in these cases
because it is conjugate to  spin chain Hamiltonians which also possess the  supersymmetry.

We note that the length-changing supersymmetry  relation for the DiSSEP Markov matrix $M^{L}$  and the supercharge $Q^{L\dag}$  for $L$ sites, 
$M^{L} Q^{L\dag}=Q^{L\dag} M^{L-1}$, is reminiscent of  a ``transfer matrix'' symmetry that has been  observed in  other exclusion processes and discuss the similarity.

\end{abstract}

\nopagebreak

\section{Lattice SUSY}

A dynamical, exact lattice supersymmetry in one dimensional lattice fermion systems and spin chains  was first observed by Fendley et.al.  \cite{Fend1,Fend2,Fend3}. A lattice Hamiltonian for $L$ sites  with such a supersymmetry can be written as 
 \be\label{HN}  H^{L}= Q^{L\dag} Q^{L}+ Q^{L+1}  Q^{L+1\dag} \label{eq:HSUSYdef}\ee
where $ H^{L}$  acts on the vector space $V^{\ot L}$, with $V\simeq \C^2$.
The lattice supercharges $Q^{L}, Q^{L\dag}$ act on chains of length $L$ and $L-1$ respectively 
as $Q^{L}: V^{\ot L}\ra V^{\ot (L-1)}$ and $Q^{L\dagger}:V^{\ot (L-1)}\ra V^{\ot L}$ \footnote{The choice of $Q^{L\dagger}$ and $\q^{\dagger}$ to be creation operators, which seems appropriate in this context, is the opposite of that used in \cite{Hag1,Hag2,Hag3}  but agrees with that in  \cite{WY}.}. 
For an open chain, these may be expressed in terms of {\em local} supercharges as 
\be
Q^{L}=\sli_{k=1}^{L-1} (-1)^{k+1} \q_{k,k+1},\quad Q^{L\dag}=\sli_{k=1}^{L} (-1)^{k+1} \q^\dag_k \label{eq:Qdef}\ee
where  $\q : V\ot V\ra V $ and $\q^\dag: V \ra  V \ot V$ and the subscripts denote the lattice sites on which the operators act \cite{Hag1}.  In a matrix representation $\q$ and $\q^\dag$ are thus $2 \times 4$ and $4 \times 2$ matrices respectively.
Satisfying the standard nilpotency conditions for the global  supercharges
 \be  Q^{L-1} Q^{L}=0,\quad Q^{L+1\dagger} Q^{L\dagger}=0,\label{eq:inilpotency}\ee
gives the following associativity condition on the local supercharge $\q$ for open chains \cite{Hag2,WY} 
\be\label{eq:assoc}
\q(\q\ot \id) = \q(\id \ot \q) 
\ee
or the equivalent coassociativity  condition on $\q^{\dagger}$
\be (\q^\dag\ot \id) \q^\dag = (\id \ot \q^\dag) \q^\dag \; .
\label{eq:coassoc}
\ee
The condition on $\q^{\dagger}$ (and similarly for $\q$) for closed chains is modified to
\be 
\left[ (\q^\dag\ot \id) \q^\dag -  (\id \ot \q^\dag) \q^\dag \right] | \psi \rangle  = | \chi \rangle \otimes | \psi  \rangle - | \psi \rangle \otimes | \chi \rangle  \, , \qquad \forall | \psi \rangle \in V
\label{eqn:nilpot}
\ee
where $| \chi \rangle \in V \otimes V$ is some fixed vector.

If a supercharge of the form eq.(\ref{eq:Qdef}) satisfying  equs.(\ref{eq:assoc},\ref{eq:coassoc}) or equ.(\ref{eqn:nilpot})  is inserted into equ.(\ref{HN}) all the non-nearest-neighbour terms in the anticommutator  cancel due to the alternating sign factors and 
the resulting nearest-neighbour bulk Hamiltonian is of the form
\begin{equation}
h = -(\id \otimes \q)(\q^\dag\otimes \id) - (\q\otimes \id)(\id \otimes \q^\dag)+\q^\dag\q + \frac12\left(\q\q^\dag \otimes \id +\id \otimes \q\q^\dag \right) \, ,
\label{eqn:HamiltonianDensity}
\end{equation}
supplemented by boundary terms  $(1/2) \q\q^\dagger$ for open chains.
Using the  nilpotency conditions  in eq.(\ref{eq:inilpotency}) shows that the supercharges relate the Hamiltonians of chains of different length, i.e.
\be H^{L-1} Q^{L}=Q^{L} H^{L},\quad  H^{L} Q^{L\dag}=Q^{L\dag} H^{L-1} \, .\label{eq:SUSY}\ee

Various  choices of $\q, \q^\dag$ leading to well-known Hamiltonians have been explored.  Fendley and Yang \cite{Fend2} noted that
\be
\q^\dag | 0 \rangle = \emptyset\quad  \q^\dag | 1 \rangle = | 00 \rangle 
\ee
or, in matrix form
\be
\q = \left[ 
\begin{matrix}  
0 &  0&  0 & 0 \\
1 &  0&  0 & 0
\end{matrix}
\right]
\ee                  
gave (up to a constant term)  the $XXZ$ Hamiltonian at its combinatorial point with diagonal boundary conditions 
\be H_{comb}=-\frac{1}{2} \sli_{k=1}^{L-1} \left(\sigma^x_k \sigma^x_{k+1} + \sigma^y_k \sigma^y_{k+1} -\frac{1}{2}\left( \sigma^z_k \sigma^z_{k+1} - \id \right) \right)-\frac{1}{4} (\sigma_1^z+\sigma_L^z).\label{eq:XXZ1}\ee
We have dropped the superscript $L$ on the Hamiltonian above, and henceforward, for notational conciseness.
Hagendorf et. al. \cite{Hag2} observed that this supercharge can be combined with its image under spin reversal ($ | 0 \rangle \to | 1\rangle $)
\be\label{reversed}
\bar\q = \left[ 
\begin{matrix}  
0 &  0&  0 & 1 \\
0 &  0&  0 & 0
\end{matrix}
\right]
\ee
and a gauge supercharge which  acts on any vector $|\psi\rangle \in V$ as
\be
\q_{\phi}^\dag | \psi \rangle = | {\phi} \rangle \ot | \psi \rangle + | \psi \rangle \ot  | {\phi} \rangle \, , 
\ee
where  $|\phi\rangle$ is some vector in $V$, to give a one parameter family of supercharges
\be\label{eq:qy}
\q (y)= x \left[ 
\begin{matrix}  
-2 y  &  -y^2 &  -y^2  & y^3 \\
1 &  -y &  -y  & -2 y^2
\end{matrix}
\right]
\ee        
with $x = ( 1 + | y|^6)^{-1/2}$. These still produced the same $XXZ$ bulk Hamiltonian when inserted into eq.(\ref{eqn:HamiltonianDensity}) but gave identical left and right, now non-diagonal, boundary terms that  depended explicitly on $y$. The supercharge
$Q^{L}$  resulting from eq.(\ref{eq:qy}) can be further elaborated to give a limited class of  non-identical boundary terms \cite{Hag2}. 
As we discuss in section \ref{sec:closed}, a  different choice of $\q$ gives a one-parameter family of $XYZ$ Hamiltonians \cite{Fend3,Hag3} for closed chains and the approach readily generalises to higher spin models \cite{Hag1,Matsui} and $\gl N  M$ Hamiltonians \cite{GLNM}. 

There is  a close relation between one-dimensional quantum spin chains  and various one-dimensional  exclusion processes, so a natural question to pose is whether the dynamical lattice supersymmetry might also exist in such models. This can be  answered  in the  affirmative for  at least one model (with a particular choice of parameters), the  Dissipative  Symmetric Simple Exclusion Process (DiSSEP), which is described in the next section.

\section{The DiSSEP}

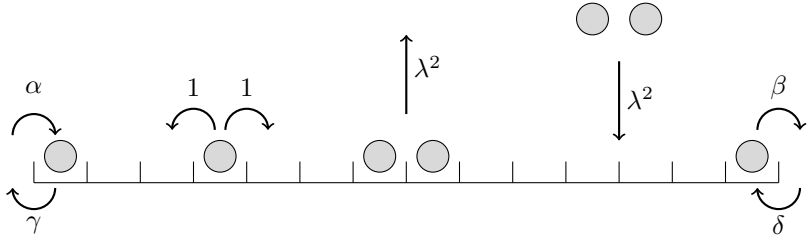
\begin{figure}[h!]
\begin{center}
 \begin{tikzpicture}[scale=0.7]
\draw (-2,0) -- (12,0) ;
\foreach \i in {-2,-1,...,12}
{\draw (\i,0) -- (\i,0.4) ;}
\draw[->,thick] (-2.4,0.9) arc (180:0:0.4) ; \node at (-2.,1.8) [] {$\alpha$};
%\draw[->,thick] (-2.4,0.9) -- (-1.6,0.9); \node at (-2.,1.8) [] {$\alpha$};
\draw[->,thick] (-1.6,-0.1) arc (0:-180:0.4) ; \node at (-2.,-0.8) [] {$\gamma$};
\draw  (-1.5,0.5) circle (0.3) [fill=gray!30,circle] {};
\draw  (1.5,0.5) circle (0.3) [fill=gray!30,circle] {};
\draw  (4.5,0.5) circle (0.3) [fill=gray!30,circle] {};
\draw  (5.5,0.5) circle (0.3) [fill=gray!30,circle] {};
\draw  (8.5,3.1) circle (0.3) [fill=gray!30,circle] {};
\draw  (9.5,3.1) circle (0.3) [fill=gray!30,circle] {};
\draw  (11.5,0.5) circle (0.3) [fill=gray!30,circle] {};
\draw[->,thick] (1.4,1) arc (0:180:0.4); \node at (1.,1.8) [] {$1$};
\draw[->,thick] (1.6,1) arc (180:0:0.4); \node at (2.,1.8) [] {$1$};
\draw[->,thick] (5,1.3) -- (5,2.8); \node at (5.4,2.2) [] {$\lambda^2$};
\draw[->,thick] (9,2.3) -- (9,0.8); \node at (9.4,1.6) [] {$\lambda^2$};
\draw[->,thick] (11.6,1) arc (180:0:0.4) ; \node at (12.,1.8) [] {$\beta$};
\draw[->,thick] (12.4,-0.1) arc (0:-180:0.4) ; \node at (12.,-0.8) [] {$\delta$};
 \end{tikzpicture}
 \end{center}
 \caption{Allowed DiSSEP moves and their rates}
 \label{fig:DiSSEP}
\end{figure}

The  DiSSEP was presented in \cite{DiSSEP}  as an integrable deformation of the  Symmetric Simple Exclusion Process (SSEP) which still allowed a solution via the matrix product  ansatz. A concise way to describe the dynamics in such systems  is to use Dirac braket notation to describe the state.  For an open system with $L$ sites, introduce an indicator variable $n_i \in \{0,1\}$ at each site $i$ to denote the presence or absence of a particle and denote the probability of finding a configuration $n_1\dots,n_{L}$ at time $t$ by  $P_t(n_1,\dots,n_{L})$. The evolution of the ket vector $|P_t\rangle$
\begin{equation}
 |P_t\rangle=\sum_{n_1,\dots,n_L\in \{0,1\}}  P_t(n_1,\dots,n_{L})~ |n_1\dots n_L\rangle \; ,
\end{equation}
where  $|n_1\dots n_L\rangle=|n_1\rangle\otimes\dots\otimes|n_L\rangle$
and the basis vectors are 
$|0\rangle=\left(
 \begin{array}{c}
 1\\0
 \end{array}
\right)$ and $|1\rangle=\left(
 \begin{array}{c}
 0\\1
 \end{array}
 \right)$,
 is given by  master equation 
\begin{equation}
 \frac{d|P_t\rangle}{dt}=M\ |P_t\rangle \, .
\end{equation}
The Markov matrix $M$ appearing in the master equation is given for the DiSSEP by
\begin{equation} \label{Markov-matrix}
 M (\lambda^2) =B_1+\sum_{k=1}^{L-1}m_{k,k+1}+\overline{B}_L
\end{equation}
with boundary transition matrices $B$, $\overline{B}$ and bulk transition matrix $m$  given by
\begin{equation} \label{eq:mBB}
B =\left( \begin {array}{cc} 
-\alpha&\gamma\\ 
\alpha&-\gamma
\end {array} \right)\quad ,\qquad  m=\left( \begin {array}{cccc} 
-\lambda^2&0&0&\lambda^2\\ 
0&-1&1&0\\
0&1&-1&0\\
\lambda^2&0&0&-\lambda^2
\end {array} \right)
\quad\text{,}\qquad
\overline{B} =\left( \begin {array}{cc} 
-\delta&\beta\\ 
\delta&-\beta
\end {array} \right)\;.
\end{equation}
The bulk Markov matrix $m_{k,k+1}$ acts between nearest neighbour sites $k, k+1$, giving  forward and backward hops and pair addition and annihilation in the bulk, while the boundary matrices $B, \overline{B}$
allow the addition and removal of particles at both ends of the system. The stochastic nature of the model is evident from the column sums of the various matrices being zero, since they describe rates. This is the distinguishing feature of this class of models. The various allowed processes for particle moves and their associated rates are shown in fig.(\ref{fig:DiSSEP}).

The Markov matrices of  such one-dimensional  exclusion processes and one-dimensional quantum spin chains 
can be related by conjugation.  
For the case of the DiSSEP, the Markov matrix $M(\lambda^2)$ is conjugate to the Hamiltonian $ H_{XXZ} (\lambda^2)$ of  an open $XXZ$ spin chain with upper diagonal boundary conditions, both with $L$ sites,  via \cite{these}
\begin{equation}\label{eq:sim}
  H_{XXZ} (\lambda^2) = - U_1U_2\dots U_L \,  M (\lambda^2) \,  U_1^{-1}U_2^{-1}\dots U_L^{-1}
\end{equation}
where
\be\label{eq:U}
 U=\begin{pmatrix} -1&1\\1&1 \end{pmatrix} \ee
and
\begin{eqnarray}
  H_{XXZ} (\lambda^2) &=& - (\alpha-\gamma)\sigma_1^+ +\frac{\alpha+\gamma}{2}(\sigma^z_1+ \id)\ -\ (\delta-\beta)\sigma_L^+ +\frac{\delta+\beta}{2}(\sigma^z_L+\id)\nonumber\\
  &&+ \frac{\lambda^2-1}{2}\ \sum_{k=1}^{L-1}
 \Big( \sigma_k^x\sigma_{k+1}^x+\sigma_k^y\sigma_{k
+1}^y-\frac{\lambda^2+1}{\lambda^2-1}(\sigma_k^z\sigma_{k+1}^z-\id) \Big)\label{eq:Hxxz}
\end{eqnarray}
with $\sigma^{x,y,z}$ being the standard Pauli matrices and   $\sigma^{+,-}$ raising and lowering matrices \footnote{We have included  minus signs in both the conjugation in eq.(\ref{eq:sim}) and the Hamiltonian in eq.(\ref{eq:Hxxz}) by comparison with \cite{these}  (in a similar manner to \cite{deGE}) to facilitate comparison with  various $H_{XXZ}$ Hamiltonians and Markov matrices later, where the natural choice is to take the minus sign in front of the Hamiltonians.}. It is clear from eq.(\ref{eq:Hxxz}) that $\lambda^2=1$ is a particularly simple, diagonal Ising limit for the bulk Hamiltonian in the model. Similarly, if $\alpha=\gamma$ and $\beta=\delta$ the $XXZ$ Hamiltonian boundary conditions also become diagonal. The simplicity is reflected in the solution of the conjugate DiSSEP when $\lambda^2=1$ \cite{DiSSEP,these}.

\section{The  open DiSSEP at $\lambda^2=1$ and Lattice SUSY}

It is straightforward to see that
\be\label{DiSSEPq}
\q = \left[ 
\begin{matrix}  
0 &  1&  1 & 0 \\
1 &  0&  0 & 1
\end{matrix}
\right]
\ee             
and its image under spin reversal
\be
\bar{\q} = \left[ 
\begin{matrix}  
1 &  0&  0 & 1 \\
0 &  1&  1 & 0
\end{matrix}
\right]\label{DiSSEPqbar}
\ee           
satisfy equs.(\ref{eq:assoc},\ref{eq:coassoc}) and that both generate the negative of the  bulk DiSSEP Markov matrix 
\be\label{Mbulk}
- m=  \left( \begin {array}{cccc} 
1 &0&0&-1\\ 
0&1&-1&0\\
0&-1&1&0\\
-1 &0&0&1
\end {array} \right) = - \left( \sigma^x \otimes \sigma^x - \id \right)
\label{mbulk}
\ee
for  $\lambda^2=1$ when employed in eq.(\ref{eqn:HamiltonianDensity}).  Inserting an overall minus into the relation between the supercharges and the Hamiltonian, now Markov matrix,  in    eq.(\ref{eq:HSUSYdef}) does not change any of the ensuing discussion, so the change in sign is immaterial for the existence of the lattice supersymmetry.
The boundary matrix $(1/2) \q\q^\dagger$, however, obtained from both of these supercharges is diagonal
\be
B = \bar B =  \left[ 
\begin{matrix}  
1 &  0 \\
0 &  1
\end{matrix}
\right]
\ee
and therefore non-stochastic.

For the $XXZ$ Hamiltonian  of  equ.(\ref{eq:XXZ1}) $\q^\dag$ and $\bar \q^\dag$ anti-commute up to boundary terms
\begin{equation}
  \label{eqn:AntiCommutationQQBar}
  \left[ (-\q^\dag \otimes 1+1\otimes \q^\dag)\bar \q^\dag+(-\bar\q^\dag \otimes 1+ 1\otimes \bar\q^\dag) \q^\dag \right] |\psi \rangle = |\chit\rangle \otimes |\psi \rangle - |\psi \rangle \otimes |\chit\rangle
\end{equation}
$\forall  \; |\psi \rangle \in V$ and where $| \chi \rangle$ is explicitly calculable, so additional gauge terms are needed to combine them into eq.(\ref{eq:qy}) to give  a $\q(y)$ that will satisfy eqs.(\ref{eq:assoc},\ref{eq:coassoc}). In the case of the DiSSEP $\q^\dag$ and $\bar \q^\dag$ from equs.(\ref{DiSSEPq},\ref{DiSSEPqbar})  anti-commute without boundary terms
\begin{equation}
  \label{eqn:AntiCommutationQQBar2}
  \left[ (-\q^\dag \otimes 1+1\otimes \q^\dag)\bar \q^\dag+(-\bar\q^\dag \otimes 1+ 1\otimes \bar\q^\dag) \q^\dag \right] |\psi \rangle = 0 \,  .
  \end{equation}
This allows them to be directly combined without introducing any gauge terms  to give a one-parameter family of supercharges $\q(y)$ which continue to satisfy
the (co)associativity conditions of eqs.(\ref{eq:assoc},\ref{eq:coassoc})
\be\label{DiSSEPqy}
\q(y) = x \left[ 
\begin{matrix}  
1 &  y&  y & 1 \\
y &  1&  1 & y
\end{matrix}
\right] \, , 
\ee        
where $x = ( 1 + |y|^2 )^{-1/2}$ in this case. When inserted into equ.(\ref{eqn:HamiltonianDensity}) $\q(y)$ still gives the (negative) DiSSEP Markov matrix 
in the bulk of eq.(\ref{mbulk})
but the boundary terms are modified to
\be
B(y) = \bar B(y) =  \left[ 
\begin{matrix}  
1 &  { 2 \Re (y) \over 1 + | y|^2}\\
{ 2 \Re (y) \over 1 + | y|^2} &  1
\end{matrix}
\right] \, .
\ee
We are thus able to obtain stochastic boundary matrices by taking $y=-1$, corresponding to the zero bias  case of 
$\alpha = \beta = \gamma = \delta =1$ when the overall minus sign is taken into account.

 The DiSSEP supercharge $\q(y)$ can be translated to its conjugate,  $\q_c (y)$, using the $U$ matrix  from equ.(\ref{eq:U})
\begin{eqnarray}
\q_c (y) &=&   U \,  \q(y) \,    U^{-1} \ot U^{-1}  \nonumber \\
\q^\dag_c (y) &=&    U \ot U \,  \q^\dag(y) \,  U^{-1}   \, ,
 \end{eqnarray}
which gives the supercharge for the $\lambda^2=1$ $XXZ$ Hamiltonian (i.e. Ising Hamiltonian $H_Z$) that is conjugate to $\lambda^2=1$ DiSSEP Markov matrix.
We find
\be
\q_c(y) = x \left[ 
\begin{matrix}  
y-1 &  0&  0 & 0 \\
0 &  0&  0 & y+1
\end{matrix}
\right]  
\label{DiSSEPqc}
\ee
and
\be
\q_c^{\dagger}(y) = 2 x \left[ 
\begin{matrix}  
\bar{y}-1 &  0\\
 0 & 0 \\
0 &  0\\
 0 & \bar{y}+1
\end{matrix}
\right]  \, ,
\label{DiSSEPqc-dagger}
\ee
where the additional, asymmetric  $2$ in $\q^\dag_c $  is due to the different factors of $U$ and $U^{-1}$ appearing in the conjugates. 
When  $\q_c(y), \q_c^\dag(y)$ are inserted into eq.(\ref{eqn:HamiltonianDensity}) they give the simple diagonal bulk and boundary  Hamiltonians
\begin{equation} \label{eq:localw8vertex}
B_c (y) = \bar B_c (y) = \left( \begin {array}{cc} 
1- {2 Re(y) \over 1 + |y|^2 }&0\\ 
0&1+{2 Re(y) \over 1 + |y|^2 }
\end {array} \right)\quad ,\qquad  m_c =\left( \begin {array}{cccc} 
0&0&0&0\\ 
0&2&0&0\\
0&0&2&0\\
0&0&0&0
\end {array} \right)
,
\end{equation}
so $m_c =  -  ( \sigma^z \ot \sigma^z - \id)$.  Consistently, this is the bulk term for $H_{XXZ}(\lambda^2 =1)$  in  eq.(\ref{eq:Hxxz}),  which is just  the Ising Hamiltonian, or $H_Z$.

The results of this section could thus equivalently be  construed as stating that  $\q_c(y)$, $\q_c^{\dagger}(y)$  of eq.(\ref{DiSSEPqc},\ref{DiSSEPqc-dagger}) provide a one parameter family of supercharges for the  diagonal Ising Hamiltonian $H_Z$
\be
 H_{Z} (y) =  - \sum_{k=1}^{L-1}
 \Big( \sigma_k^z\sigma_{k+1}^z-\id  \Big)\label{eq:Hzz} + B_{c,1}(y) + \bar B_{c,L}(y) \, .
\ee     
This Hamiltonian is conjugate to the (negative of the)  $\lambda^2=1$ DiSSEP Markov matrix, $H_{X}(y)$, generated by supercharge $\q(y)$,$\q^{\dagger}(y)$
\be
 H_{X} (y) =  - \sum_{k=1}^{L-1}
 \Big( \sigma_k^x\sigma_{k+1}^x-\id  \Big)\label{eq:Hzz} + B_{1}(y) + \bar B_{L}(y) \, ,
\ee     
via
\begin{equation}\label{eq:HZsim}
  H_{Z}(y) =  U_1U_2\dots U_L \,  H_X (y) \,  U_1^{-1}U_2^{-1}\dots U_L^{-1}
\end{equation}
(since $H_X(y)=-M$) and both therefore display the supersymmetry.  When $y=-1$ the   boundary terms  $B_1(-1)$ and $ \bar B_L(-1)$ in $H_{X}(-1)$ are stochastic 
\be
B_1(-1) = \bar B_L(-1) =  \left( 
\begin{matrix}  
1 &  -1\\
-1  &  1
\end{matrix}
\right) \, .
\ee
 These stochastic boundary terms are conjugate to 
\begin{equation} \label{eq:localw8vertex}
B_{c,1} (-1) = \bar B_{c,L} (-1) = \left( \begin {array}{cc} 
2&0\\ 
0&0
\end {array} \right) = \sigma_z+\id \, ,
\end{equation}
which can be seen to be the boundary terms  in eq.(\ref{eq:Hxxz}) when $\alpha=\beta=\delta=\gamma=1$.

\section{The closed  DiSSEP at $\lambda^2 = 1$ and Lattice SUSY}\label {sec:closed}

For a closed spin chain or a closed exclusion process, we can apply the coassociativity condition of eq.(\ref{eqn:nilpot}) with a non-zero right hand side  to sift out candidate local supercharges. One such example is the supercharge for the $XYZ$ Hamiltonian given in \cite{Hag1}
\be
\q_{XYZ}(\zeta) = \left[ 
\begin{matrix}  
0 &  0&  0 & 0 \\
1 &  0&  0 & -\zeta
\end{matrix}
\right]
\ee
i.e.                  
\be
\q_{XYZ}(\zeta)^\dag | 0 \rangle = \emptyset\quad  \q_{XYZ}(\zeta)^\dag | 1 \rangle = | 00 \rangle  - \zeta | 11 \rangle
\ee
which generates the bulk Hamiltonian
\begin{equation} \label{eq:XYZ}
H_{XYZ}(\zeta)=\left( \begin {array}{cccc} 
1&0&0&-\zeta\\ 
0&\frac{1}{2} + \frac{\zeta^2}{2}&-1&0\\
0&-1&\frac{1}{2} + \frac{\zeta^2}{2}&0\\
-\zeta&0&0&1
\end {array} \right)
\end{equation}
arising from a one-parameter family of (closed) $XYZ$ models
\be H_{XYZ}(\zeta) =-\frac{1}{2} \sli_{k=1}^{L} \left( ( 1 + \zeta) \sigma^x_k \sigma^x_{k+1} + ( 1 - \zeta)  \sigma^y_k \sigma^y_{k+1} +\frac{\zeta^2 -1}{2}\left( \sigma^z_k \sigma^z_{k+1} - \id \right) \right) +  L  \, \id  \; .\ee
$H_{XYZ}(1)$ is the negative of the $\lambda^2=1$ DiSSEP Markov matrix, i.e 
\be
H_{XYZ}(1) = 
\left(
 \begin {array}{cccc} 
1&0&0&-1\\ 
0&1&-1&0\\
0&-1&1&0\\
-1&0&0&1
\end {array} 
\right)
= -  \left( \sigma^x \otimes  \sigma^x    - \id \right)
\ee
though in this case the supercharge satisfies eq.(\ref{eqn:nilpot}) rather than eqs.(\ref{eq:assoc},\ref{eq:coassoc}), so we have
\begin{eqnarray}
  \label{eqn:AntiCommutationXYZ}
 (\q_{XYZ}(1)^\dag \otimes \id- \id \otimes \q_{XYZ}(1)^\dag) \q_{XYZ}(1)^\dag   |\psi \rangle  
  = -|00\rangle \otimes |\psi\rangle + |\psi\rangle \otimes |00\rangle
\end{eqnarray}
$\forall  \; |\psi\rangle \in V$, i.e. $| \chit \rangle = - | 00 \rangle$ in  eq.(\ref{eqn:nilpot}). 

It is therefore possible  in a closed system
 for different $\q$'s, in this case
\begin{eqnarray}
\q(y) = x \left[ 
\begin{matrix}  
1 &  y&  y & 1 \\
y &  1&  1 & y
\end{matrix}
\right] \nonumber \\
\q_{XYZ}(1) = \left[ 
\begin{matrix}  
0 &  0&  0 & 0 \\
1 &  0&  0 & -1
\end{matrix} 
\right] \, ,
\end{eqnarray}
to produce the same bulk DiSSEP Markov matrix, $ -  \left( \sigma^x \otimes  \sigma^x    - \id \right)$.

\section{The open  DiSSEP at $\lambda^2 = - 3$ and Lattice SUSY}

The exact dynamical lattice supersymmetry also exists in the open DiSSEP  at the unphysical value of  $\lambda^2 = -3$,  since the conjugate Hamiltonian in this case is a multiple of the $XXZ$ Hamiltonian at its combinatorial point, which possesses the supersymmetry. 

If we define
 \be
\label{DiSSEPq3}
\hat \q 
=   {1 \over \sqrt{2}} \left[ 
\begin{matrix}  
-1  &  -1 &  -1  & -1 \\
1 &  1 &  1  & 1
\end{matrix}
\right] 
\ee        
the (co)associativity  conditions equ.(\ref{eq:assoc},\ref{eq:coassoc}) are  satisfied and the 
corresponding bulk Markov matrix obtained  from  eq.(\ref{eqn:HamiltonianDensity}) is 
\be\label{Mbulk}
m=   \left( \begin {array}{cccc} 
1 &0&0&3\\ 
0&5&-1&0\\
0&-1&5&0\\
3 &0&0& 1
\end {array} \right) 
= 
-   \left( \begin {array}{cccc} 
3 &0&0&-3\\ 
0&-1 &1&0\\
0&1&-1&0\\
-3 &0&0& 3
\end {array} \right)  + 4  \id
\ee
which is minus the DiSSEP Markov matrix at $\lambda^2=-3$ along with a constant term,  together with stochastic boundary matrices $(1/2) \hat \q \hat \q^\dagger$
\be
B = \bar B =  \left[ 
\begin{matrix}  
1 &  -1 \\
-1 &  1
\end{matrix}
\right] \, .
\ee
When $\lambda^2=-3$ the bulk $XXZ$ Hamiltonian conjugate to the DiSSEP 
\begin{eqnarray}
  H_{XXZ} (\lambda^2)=\frac{\lambda^2-1}{2}\ \sum_{k=1}^{L-1}
 \Big( \sigma_k^x\sigma_{k+1}^x+\sigma_k^y\sigma_{k+1}^y-\frac{\lambda^2+1}{\lambda^2-1}(\sigma_k^z\sigma_{k+1}^z-\id) \Big)\label{eq:Hxxzbulk}
\end{eqnarray}
is four times   the $XXZ$ Hamiltonian at its combinatorial point, $H_{comb}$,  in eq.(\ref{eq:XXZ1}), i.e.
\begin{eqnarray}
  H_{XXZ} (-3) = 4 H_{comb} =-2 \ \sum_{k=1}^{L-1}
 \Big( \sigma_k^x\sigma_{k+1}^x+\sigma_k^y\sigma_{k+1}^y-\frac{1}{2}(\sigma_k^z\sigma_{k+1}^z-\id) \Big)  \,  . \label{eq:Hxxzbulk2}
\end{eqnarray}
On the other hand, the conjugates of the supercharge $\hat \q$ and $\hat \q^{\dagger}$ from eq.(\ref{DiSSEPq3})  which give  the $\lambda^2=-3$ DiSSEP are
\be
\hat \q_c= U \, \hat \q \, U^{-1}  \ot U^{-1}  =   \sqrt{2} 
\left[ 
\begin{matrix}  
0&  0&  0 & 1 \\
0 &  0&  0 & 0
\end{matrix}
\right] \, .
\ee
and
\be
\q^\dag_c  &=&    U \ot U \,  \q^\dag \,  U^{-1}   \, = 2   \sqrt{2} 
\left[ 
\begin{matrix}  
0&  0\\
0 & 0 \\
0 &  0\\ 
1 & 0
\end{matrix}
\right] \, .
\ee 
$\hat \q_c$ , $\hat \q_c^{\dagger} $ are multiples of the spin reversed supercharge $\bar \q$,  $\bar \q^{\dagger} $  for $H_{comb}$ in eq.(\ref{reversed}), so substituting  them into eq.(\ref{eqn:HamiltonianDensity})
gives  $4 H_{comb}$, consistently with eq.(\ref{eq:Hxxzbulk2})

Thus, just as for the $\lambda^2=1$ DiSSEP, the supersymmetry observed in the   $\lambda^2=-3$ DiSSEP is a consequence of the Markov matrix being conjugate to a  spin chain Hamiltonian which displays the supersymmetry.

\section{Conclusions}

A brute force scan by computer  of possible integer entries $\{ \ldots  \pm 2 , \pm 1, 0 , \pm1, \pm 2  \ldots \}$ in $\q$ reveals that while it is relatively easy to generate solutions of eq.(\ref{eq:assoc},\ref{eq:coassoc}), demanding that these should represent bulk stochastic matrices (column sum zero, up to a possible constant term)  and that the boundary matrices also be stochastic leaves only the two
open DiSSEP  cases discussed here, $\lambda^2=1$ and the unphysical value of $\lambda^2=-3$. As we have noted, the Markov matrices for these are conjugate to a diagonal Hamiltonian and the $XXZ$ Hamiltonian at its combinatorial point respectively, in both cases with diagonal boundary conditions.

Supersymmetric Hamiltonians/Markov matrices  for a closed system  are considered only briefly here. Since the $XYZ$ Hamiltonian at $\zeta=1$ is identical to the (negative) $\lambda^2=1$ DiSSEP Markov matrix,  two different supercharges can produce the same bulk Markov Matrix/Hamiltonian. 
As suggested in \cite{Hag1}, the classification of possible supersymmetric Hamiltonians up to equivalence under conjugations would be an interesting exercise, but is beyond the scope of this paper. We have  made no attempt to explore  conjugations and equivalences systematically along the lines of \cite{Malte} and it is possible that other open and closed stochastic Markov matrices  might be accessible from  known supersymmetric Hamiltonians using such methods. 

\begin{figure}[h!]
\begin{center}
 \begin{tikzpicture}[scale=0.6]
\draw (-2,0) -- (12,0) ;
\foreach \i in {-2,-1,...,12}
{\draw (\i,0) -- (\i,0.4) ;}
\draw[->,thick] (-2.4,0.9) arc (180:0:0.4) ; \node at (-2.,1.8) [] {$\alpha$};
%\draw[->,thick] (-2.4,0.9) -- (-1.6,0.9); \node at (-2.,1.8) [] {$\alpha$};
%\draw[->,thick] (-1.6,-0.1) arc (0:-180:0.4) ; \node at (-2.,-0.8) [] {$\gamma$};
\draw  (-1.5,0.5) circle (0.3) [fill=gray!30,circle] {};
\draw  (1.5,0.5) circle (0.3) [fill=gray!30,circle] {};
\draw  (5.5,0.5) circle (0.3) [fill=gray!30,circle] {};
\draw  (6.5,0.5) circle (0.3) [fill=gray!30,circle] {};
\draw  (11.5,0.5) circle (0.3) [fill=gray!30,circle] {};
%\draw  (9.5,3.1) circle (0.3) [fill=gray!30,circle] {};
%\draw  (10.5,3.1) circle (0.3) [fill=gray!30,circle] {};
%\draw[->,thick] (1.4,1) arc (0:180:0.4); \node at (1.,1.8) [] {$1$};
\draw[->,thick] (1.6,1) arc (180:0:0.4); \node at (2.,1.8) [] {$1$};
\draw[->,thick] (6,1.3) -- (6,2.8); \node at (6.4,2.2) [] {$\lambda^2$};
%\draw[->,thick] (10,2.3) -- (10,0.8); \node at (10.4,1.6) [] {$\lambda^2$};
\draw[->,thick] (11.6,1) arc (180:0:0.4) ; \node at (12.,1.8) [] {$\beta$};
%\draw[->,thick] (12.4,-0.1) arc (0:-180:0.4) ; \node at (12.,-0.8) [] {$\delta$};
 \end{tikzpicture}
 \end{center}
 \caption{Allowed ASAP moves and rates}
 \label{fig:ASAP}
\end{figure}
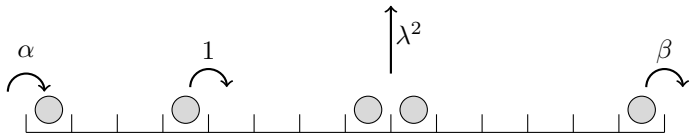

The investigations here were originally motivated by the observation that a ``transfer matrix'' symmetry which takes the form
\be\label{eq:trans}
M^{L} T^{L\dag}=T^{L\dag} M^{L-1} 
\ee
exists in  several stochastic models, which is analogous to the length changing SUSY relation of eq.(\ref{eq:SUSY}).
$T^{L\dag}$ was  explicitly presented via a recursion relation for  the asymmetric annihilation process (ASAP),
whose bulk and boundary Markov matrices are given by
\begin{equation} \label{eq:mBB2}
B =\left( \begin {array}{cc} 
-\alpha&0\\ 
\alpha&0
\end {array} \right)\quad ,\qquad  m=\left( \begin {array}{cccc} 
0&0&0&\lambda^2\\ 
0&0&1&0\\
0&0&-1&0\\
0&0&0&-\lambda^2
\end {array} \right)
\quad\text{,}\qquad
\overline{B} =\left( \begin {array}{cc} 
0&\beta\\ 
0&-\beta
\end {array} \right)\;.
\end{equation}
 in \cite{ASAP}.
The allowed moves for the ASAP are  shown in fig.(\ref{fig:ASAP}). While it is tempting to regard the transfer matrix symmetry as evidence for a similar dynamical lattice supersymmetry to the one discussed here for the DiSSEP,  the bulk Markov matrix in eq.(\ref{eq:mBB2}) was not amongst those  generated by scanning through various potential $\q$'s here. The algorithm for determining  $T^{L\dag}$ in \cite{ASAP} is based on the recursive properties of the Markov matrix and is a global construction rather than a local formulation, giving no indications of nilpotency for $T^{L\dag}$.
A further point of divergence is that the transfer matrix symmetry exists for generic $\alpha, \beta$ in the ASAP whereas demanding dynamical lattice supersymmetry in the DiSSEP along with stochastic  boundaries  constrains $\alpha=\beta=\gamma=\delta=1$ and  $\lambda^2=1$ (or $\lambda^2=-3$ if we allow unphysical values).

 A similar situation exists  for the Totally Asymmetric Exclusion Process (TASEP)  \cite{TASEP}. For this a relation between the Markov matrices for  systems of different lengths is of the form 
\be
M^{L} \tilde T^{L\dag}=T^{L\dag} M^{L-1}  \; ,
\label{eq:TTbar}
\ee
where $\tilde T^{L\dag}$ and $T^{L\dag}$ are now two different matrices. Again, the Markov matrix for the TASEP
\be
m=\left( \begin {array}{cccc} 
0&0&0&0\\ 
0&0&1&0\\
0&0&-1&0\\
0&0&0&0
\end {array} \right)
\ee
is not produced by the class of $\q$'s we have examined.

In summary, we have shown that the open  DiSSEP  possesses a dynamical lattice supersymmetry in the sense of  \cite{Fend1,Fend2,Fend3,Hag1,Hag2,Hag3}  for $\lambda^2=1,-3$ and $\alpha=\beta=\gamma=\delta=1$.  Both the boundary conditions, which give no driving current in the DiSSEP, and the bulk Markov matrices represent particular simplifying values for the model parameters. The bulk Markov matrices for  $\lambda^2=1,-3$  are conjugate to a diagonal Ising  Hamiltonian and an $XXZ$ Hamiltonian at its combinatorial point respectively, which are themselves supersymmetric. 

While the formal similarity between the length changing supersymmetry  for various spin chains in eq.(\ref{eq:SUSY}) and the global transfer matrix symmetry in eqs.(\ref{eq:trans},\ref{eq:TTbar}) in  the ASAP \cite{ASAP} and TASEP \cite{TASEP}  is intriguing, this does not seem to be the consequence of a similar dynamical lattice supersymmetry with local supercharges in  the latter models.

\subsection*{Acknowledgements}
DAJ  would like to thank Robert Weston and Junye Yang for useful discussions. This work was supported by EPSRC grant EP/R009465/1.

\newpage

%\bibliography{susy}

\end{document}